\documentclass[aip,reprint,amsmath,amssymb,superscriptaddress]{revtex4-1}
\usepackage{graphicx}
\usepackage{graphics}
\usepackage{bbm}
\usepackage{amsmath,amsfonts,amssymb}
\usepackage{wrapfig}

\begin{document}
\title{Suppressing On-Chip EM Crosstalk for Spin Qubit Devices}
\author{S. Blanvillain}
\affiliation{ARC Centre of Excellence for Engineered Quantum Systems, School of Physics, The University of Sydney, Sydney, NSW 2006, Australia.} 
\author{J. I. Colless}
\affiliation{ARC Centre of Excellence for Engineered Quantum Systems, School of Physics, The University of Sydney, Sydney, NSW 2006, Australia.} 
\author{D. J. Reilly$^*$}
\affiliation{ARC Centre of Excellence for Engineered Quantum Systems, School of Physics, The University of Sydney, Sydney, NSW 2006, Australia.} 
\author{H. Lu}
\affiliation{Materials Department, University of California, Santa Barbara, California 93106, USA.}
\author{A. C. Gossard}
\affiliation{Materials Department, University of California, Santa Barbara, California 93106, USA.}
\begin{abstract}
We report the development and performance of on-chip interconnects designed to suppress electromagnetic (EM) crosstalk in spin qubit device architectures with the large number of gate electrodes needed for multi-qubit operation. Our design improves the performance of typical device interconnects via the use of miniaturised ohmic contacts and interspersed ground guards. Low temperature measurements and numerical simulation confirm that control and readout signal crosstalk can be suppressed to levels of order 1\%, from dc to 1 GHz. 
\end{abstract}

\maketitle
Implementing quantum error correction (QEC) in the laboratory is a formidable challenge because it is contingent on realising quantum hardware with extremely low base error thresholds \cite{Preskill:1998vu}. Recent approaches to QEC using so-called surface codes \cite{surface} suggest a far less stringent error threshold may be within reach \cite{Loss}, spurring efforts to develop multi-qubit devices that can enable fault tolerant algorithms \cite{Reed_Nature}. Crosstalk is a key issue for hardware performance \cite{Martinis_wirebond,Colless}, reducing the fidelity of qubit control and readout, as well as impacting QEC by producing non-Markovian noise  and correlated errors that can lead to simultaneous faults on otherwise independent qubits \cite{Preskill:1998vu,Preskill_cor,Knill}. 

In this paper we focus on signal fidelity and crosstalk in qubit architectures \cite{Loss:1998via,Hanson:2007eg} that comprise single electron spin-states, confined and controlled electrically in a semiconductor heterostructure via radio frequency waveforms applied to surface electrodes and transmission lines. These waveform signals control single- and two-qubit operations by modulating the exchange interaction between electrons using nanosecond voltage pulses, or by creating resonant magnetic fields at the spin transition frequency \cite{Koppens:2006kz,Kane:1998wha,Nowack:2007du}. Crosstalk in these devices is largely due to unmitigated electromagnetic (EM) coupling between control and, or, readout channels, potentially degrading the performance of single- and multi-qubit systems. 
\begin{figure*} \centering
\includegraphics[scale=0.8]{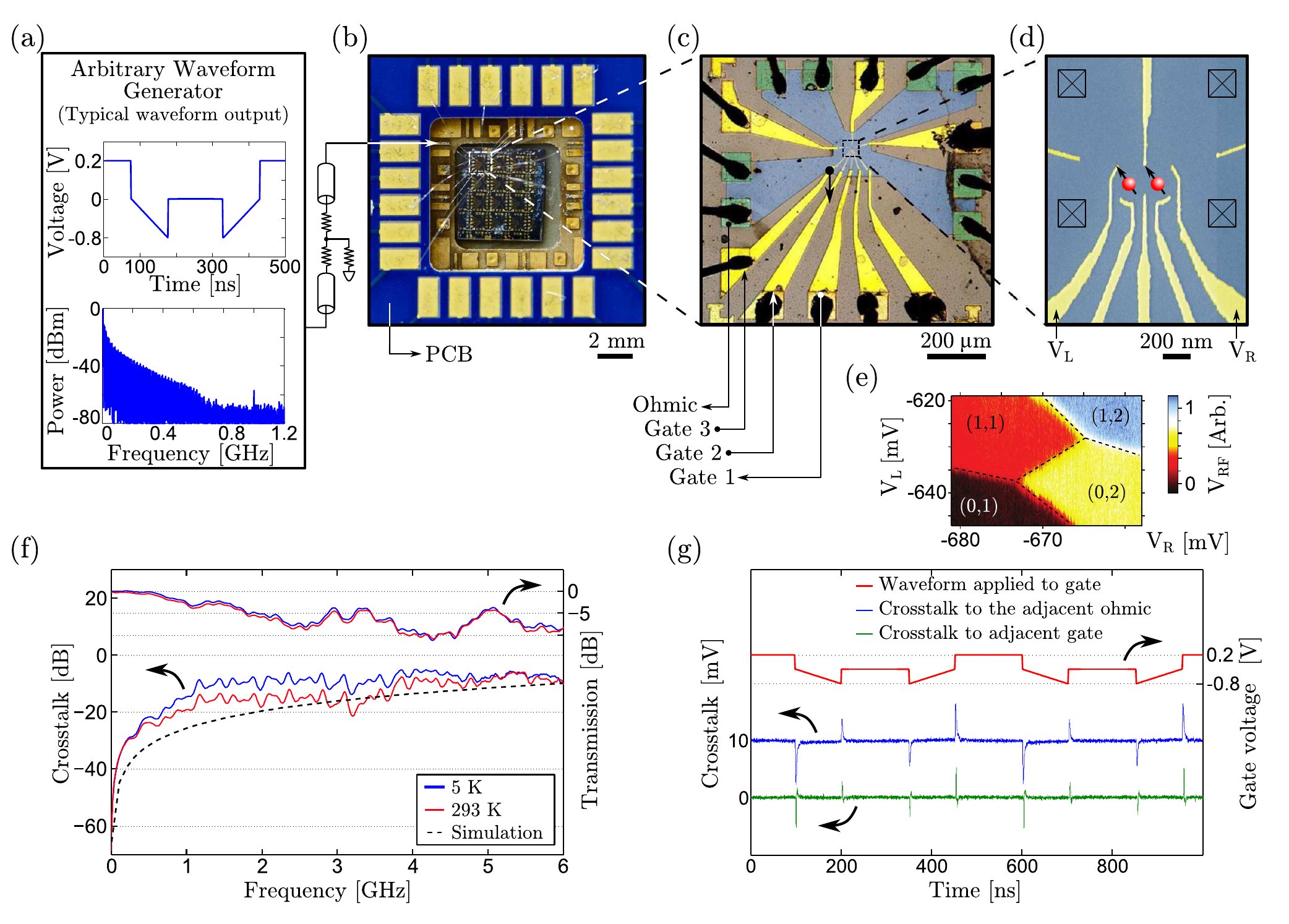}
\caption{\label{fig:photos} (Color online)  (a)  Typical time-domain waveform used for control of singlet-triplet spin qubits, with corresponding frequency components. (b) Photograph of the qubit chip linked to a multilayer PCB with Al bond wires. Attenuated coaxial cables link the room temperature arbitrary waveform generator to the PCB. (c) Optical micrograph of the device (false colour) showing bond wires (black), gate interconnects (yellow), and ohmic contacts (green). The mesa structure that defines the 2DEG is shaded blue. (d) Scanning electron micrograph (false colour)  of a representative device, showing EBL fine gates and the location of the two quantum dots and electron spins. Crossed boxes indicate regions connected to ohmic contacts. (e) Charge sensing data taken at $T$ = 20 mK on the device shown in (d). The data show the gate voltage parameter space for a two-electron (singlet-triplet) qubit indicating the magnitude of the gate voltages used to control the qubit energy levels or change the number of electrons in the double quantum dot (V$_{\mathrm{L}}$ and V$_{\mathrm{R}}$ are voltages applied to the outer gates in Fig. 1(d)). Colour axis is the readout detector response based on a rf-quantum point contact \cite{Reilly:2008ib}. Labels (m,n) indicate the number of electrons occupying the left and right quantum dots. (f) Measured and simulated crosstalk between interconnects for gates 1 and 2 and transmission along the interconnect for gate 3. (g) Low temperature ($T$ = 5 K) time-domain crosstalk between the interconnects for gates 1 and 2 (green) and between gate 3 and the adjacent ohmic (blue).}
\end{figure*}

\begin{figure} \centering
\includegraphics[width=\linewidth]{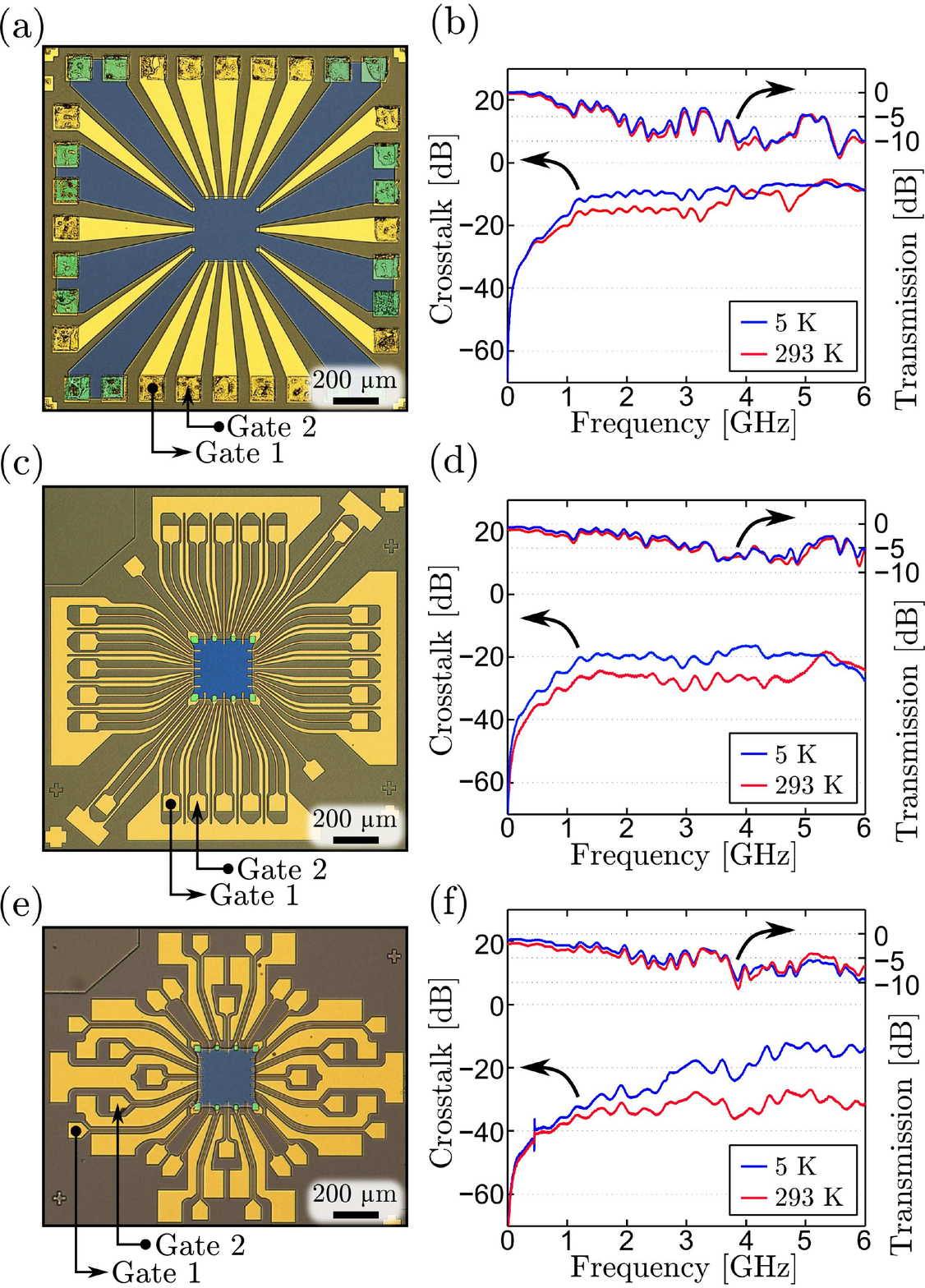}
\caption{(Color online) (a,c,e) Three generations of interconnect pattern and associated crosstalk performance. Devices do not feature fine EBL-defined gates. Gate interconnect metallisation is shown in yellow, ohmics contacts in green, with the 2DEG region shaded blue. (b,d,f) show crosstalk and transmission performance for each interconnect pattern. Crosstalk is measured between gate interconnects 1 and 2, with transmission evaluated along the length of gate interconnect 1.}
\label{fig2}
\end{figure}

We report the development and characterisation of on-chip interconnects that improve control fidelity and suppress signal crosstalk for devices with the large number of tightly-packed gate electrodes needed for multi-qubit operation. In comparison to previously used spin qubit interconnect architectures \cite{Petta:2005kn,Reilly:2008ib,Reilly:2010gp}, we show via low temperature measurements and numerical simulation, that EM crosstalk can be suppressed to 1\% levels, an improvement of more than an order-of-magnitude.   

For spin qubits based on two electron spin-states, technical improvements in control pulse transmission have recently been shown to extend coherence \cite{Bluhm}. Further approaches to advancing qubit control include clever pulse-shaping and the use of additional anti-phase signals that null crosstalk. These however, add a complexity burden and, in the case of controlling multi-qubit devices, lead to stringent clocking and qubit synchronization overheads. Motivated by the desire to forgo these complications, we have focused efforts to improve the base performance of spin qubit control hardware, firstly at the circuit board level \cite{Colless}, and here at the level of on-chip interconnects.  

A typical pulse waveform used to control (singlet-triplet) spin qubits is shown in Fig. 1(a) \cite{Petta:2005kn}. Most of the waveform power is in frequency components that span a bandwidth of several hundred MHz, although qubit control using spin resonance methods typically make use of narrow-band signals in the GHz \cite{Koppens:2006kz,Nowack:2007du}. In a typical setup, these control signals are generated at room temperature \cite{tektronix5014} and transmitted to the device at cryogenic temperatures using highly attenuated coaxial lines. In our setup a low crosstalk multilayer printed circuit board (PCB) \cite{Colless} is used at milli-Kelvin temperatures to connect coaxial cables to the qubit chip via Al bond wires, as shown in Fig. 1(b). Crosstalk at the PCB level is measured \cite{Colless} to be at most -70 dB in the frequency range of interest and is neglected here.
\begin{figure*}
\centering
\includegraphics[scale=0.8]{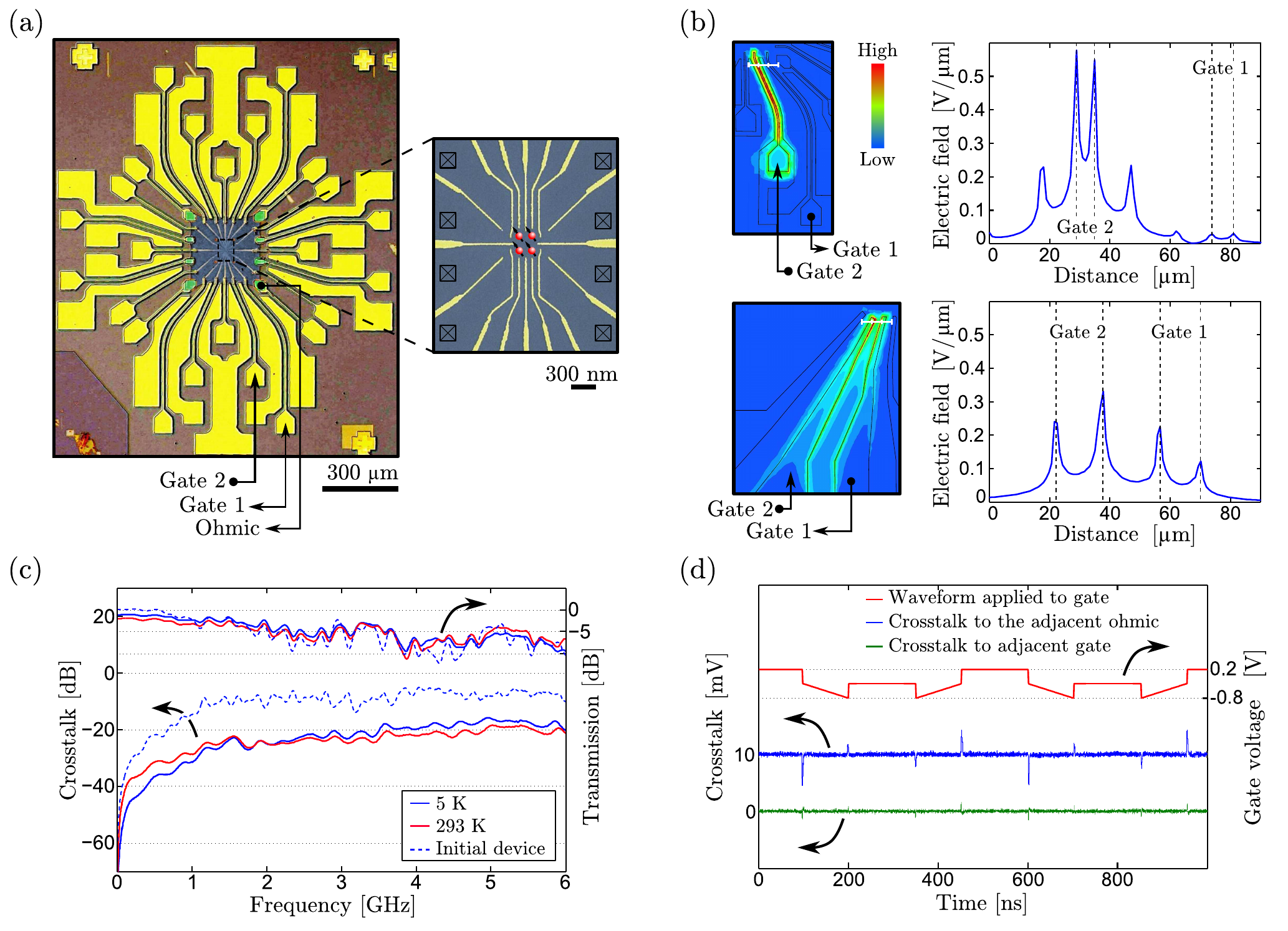}
\caption{(Color online) Characterization of an integrated low-crosstalk four dot device. (a) Optical photograph and (false colour) electron micrograph of the device. Contact gates and ground guards are shaded yellow, with ohmic contacts shaded green and 2DEG region in blue.  (b) Comparison of the simulated electric field (E-field) induced by gate interconnect 2 on gate interconnect 1 for this device (upper panels) and the original device considered in Fig. 1(c) (lower panels). Electric fields are plotted for the slice indicated by the white line in the colour images for a signal frequency of 6 GHz. (c) Crosstalk between gate interconnects 1 and 2 and transmission performance comparing this device with the original interconnect geometry considered in Fig. 1. (d) Time-domain crosstalk between gate interconnect 1 and 2 (green) and between gate 1 and the adjacent ohmic (blue) for the device shown in Fig. 3(a).} \label{fig3}
\end{figure*}

The qubit devices investigated comprise Ti-Au metal electrodes deposited on the surface of a GaAs/Al$_{0.3}$Ga$_{0.7}$As heterostructure with two dimensional electron gas (2DEG) 110 nm below the surface and low temperature mobility of $\sim$ 440,000 cm$^2$/Vs  and electron density of 2.4 $\times$ 10$^{11}$ cm$^{-2}$. Gate electrodes are defined using electron beam lithography (EBL), (see Fig. 1(d)) and contact larger interconnect metallisation defined using standard optical lithographic methods (see Fig. 1(c)). This interconnect metallisation provides a link from the on-chip bond pads to the fine EBL-defined gate electrodes and is the dominant source of on-chip EM crosstalk. Ohmic contacts to the 2DEG are produced by evaporation of a NiAuGe stack and subsequent thermal anneal, yielding typical dc resistances of order 100 $\Omega$. At GHz frequencies the impedance of the contacts is reduced due to the self capacitance of the contact metallisation.

Control of singlet-triplet qubits is performed by rapidly varying the energy levels of the confined electrons, modulating the charging and exchange energy via voltage pulses applied to the surface gates. The typical amplitude range for these pulses can be taken from Fig. 1(e), which shows the response of a rf quantum point contact \cite{Reilly:2008ib} readout detector as a function of the gate voltages that control the qubit energy levels. The data, taken at $T$ = 20 mK on the device shown in Fig. 1(d), illustrate that gate voltage changes on the order of $\pm$ 20 mV span the two-electron energy space. 

We first investigate the on-chip transmission and crosstalk performance of the typical interconnect pattern, previously used for controlling spin qubits \cite{Petta:2005kn,Barthel:2009hx} and shown in Fig. 1(c,d). This  pattern comprises 8 ohmic contacts (4 independent pairs) to the 2DEG (green squares in Fig.1(c)) and 8 gate interconnects that link wire bonds from the PCB to EBL-defined gate electrodes. We perform crosstalk characterisation at room temperature and 5 K \cite{Lakeshore} to determine the effect of the 2DEG which forms at low temperature in these devices.  

Crosstalk between gate 1 and gate 2 (see Fig. 1(c)) is measured in the frequency domain using a calibrated network analyzer \cite{PNA}, as shown in Fig. 1(f). We find an increase in the high frequency crosstalk of $\sim$ 7 dB at low temperature, likely the result of the increased capacitance between interconnects with 2DEG formation. Cooling the sample will also lead to increased crosstalk and transmission performance as the resistance in the gate metallisation is decreased. To determine the mechanism for the crosstalk between gate electrodes, we simulate the interconnect geometry using a finite element EM solver \cite{HFSSQ3D} and compare to measured $S$-parameters, as shown in Fig. 1(f). The numerical simulation does not account for bond wires, fine EBL-defined gate structures or 2DEG, which likely contribute the few dB of extra crosstalk seen in the measurements. The simulation confirms that the interconnect metallisation is the dominant source of on-chip crosstalk. To evaluated  the in-line performance of the gate interconnects, we measure the transmitted power as a function of frequency along the length of a electrode by bonding each end of gate 3 to pads on the PCB (see arrows on gate 3 in Fig. 1(c)). This method does not account for the loss from an additional bond wire and thus slightly over estimates the attenuation of the on-chip metal interconnect.

Low temperature time domain crosstalk is measured by applying a typical qubit control waveform to gate 3, or gate 2, and detecting signals\cite{scope} induced on the adjacent gate 1 electrode and ohmic contact (see Fig. 1(c)). We find that the coupling is predominantly capacitive, with resultant crosstalk signals resembling the expected waveforms for a high-pass filter or time-domain differentiator, as shown in Fig. 1(g). The magnitude of crosstalk is found to be similar for both adjacent gates and ohmic contacts and is of the order of 1\%. Although this level of crosstalk is relatively small, we note that an increase in the density of gates and contacts to accommodate controlling multi-qubit devices will necessitate higher coupling.

To investigate crosstalk in geometries with a larger number of gates, we first test a design that keeps the density of interconnects constant by scaling-up the size of the chip area. Figure 2(a) shows a photograph of an interconnect pattern with double the number of gates (16) and ohmic contacts (16) in comparison to the pattern shown in Fig. 1(c). Here the pattern area is increased by a factor of 1.5 to keep the distance (and thus capacitance) between interconnect structures of similar order to the initial 8 gate device. By increasing the area over which the interconnects lay, crosstalk is kept to levels similar to the previous measurements, as show in Fig. 2(b). We note that the larger area device does not contain fine EBL gates which likely increase crosstalk by a few dB. We find that the transmission performance of gate electrodes is also comparable to the smaller pattern investigated in Fig. 1.

In order to suppress crosstalk below these values, without further up-scaling of the chip area, we have developed the interconnect pattern shown in Fig. 2(c). This pattern introduces grounded guard shields that enclose each gate interconnect, essentially forming a co-planar waveguide \cite{Bronckers:2010vv}. The guards terminate electric field lines from the adjacent signal tracks and reduce nearest-neighbour EM crosstalk. In the previous design, the presence of the conducting 2DEG layer at low temperature creates additional coupling between interconnects, and to a larger degree, to the ohmic contacts. To reduce this 2DEG mediated crosstalk we have miniaturised the ``mesa" region that defines the 2DEG, including a significant down scaling in the area of ohmic contacts. We employ two different sized ohmic contacts, with areas $\sim$ 1400 $\mu$m$^2$ and  $\sim$ 780 $\mu$m$^2$. Even with the reduction in size, a low contact resistance of order 65 $\Omega$ is maintained. The design also aims to suppress crosstalk by reducing the width of the central signal track, thereby reducing capacitive coupling to proximal metallisation. Finally, we have chosen the co-planar waveguide geometry to closely match the impedance of the bonding terminals on the PCB. This improved impedance matching slightly increases the transmission performance of the gate interconnects. 

In comparison to the geometry shown in Fig. 2(a), the design in Fig. 2(c) improves the crosstalk performance mostly at higher frequencies above 1 GHz, reducing nearest neighbour interconnect coupling to below -20 dB. For qubit control using high frequency spin resonance, this improvement will assist in selectively addressing spins for programmed rotation. For two-electron (singlet-triplet) qubits that are controlled largely by signal frequencies in the dc - 1 GHz band, the design in Fig. 2(c) offers only a modest improvement in crosstalk performance. To address this shortfall we have further refined the design to include spatially staggered interconnects, as shown in Fig. 2(e). In addition, the interconnects are partitioned into those that carry dc-voltage signals (used to confine electrons) and those that transmit fast, high-frequency signals (used for control). Relative to the dc lines, bond-wire pads for high-frequency lines are located closer to the edge of chip to facilitate short bond-wires to the PCB. This improves transmission and crosstalk performance by minimising stray capacitance and controlling the impedance of the interconnect.  

To fully benchmark these on-chip interconnects we have fabricated a multi-qubit device with EBL-patterned gates that define 4 quantum dots in a square geometry, as shown in Fig. 3(a). We have also investigated devices with 4 quantum dots in a line with similar results. The presence of the EBL-defined gates is found to have only a minor effect on crosstalk performance, in part because in this design the ground-guard metallisation is positioned to overlap the mesa structure, capacitively grounding the 2DEG at high frequencies and reducing coupling between fine gates.

To compare the performance of the device in Fig. 3(a) to the original device (Fig. 1(c)), we numerically simulate the electric fields induced by high frequency signals on neighbouring interconnects, as shown in Fig. 3(b). The revised interconnect geometry (upper panel) concentrates the E-field tightly between the signal track and ground guards, with minimal capacitive crosstalk induced on proximal gates. This reduction in capacitive coupling leads to a suppression of crosstalk relative to the original device of $\sim$ 10 - 15 dB across the 6 GHz bandwidth, as illustrated by the frequency domain measurements shown in Fig. 3(c). We note that in addition to suppressing crosstalk, the design has doubled the number of gate and contact interconnects.  We again evaluate crosstalk in the time-domain by measuring the coupling of short rise-time signals between gates and ohmic contacts, as shown in Fig. 3(d). The biggest improvement of the new design is in decoupling adjacent gate electrodes to $\sim$ 0.1\% levels (green trace in Fig. 3(d)).   

We now turn to discuss how the EM crosstalk we have characterised affects the performance of single- and two-qubit quantum gates. For qubits based on singlet-triplet spin-states, changes in gate voltage by control pulses at the chip are in the range 1 - 20 mV, as indicated by the range of gate voltages spanned in Fig. 1(e). Crosstalk at the amplitudes measured in Fig. 1(g) will lead to induced voltage spurs on neighbouring gate electrodes at the 100 $\mu$V level. We note that at dilution fridge base temperatures these signals are many orders of magnitude above the intrinsic thermal noise background (which is well below nV/$\sqrt{\mathrm{Hz}}$ levels). To what extent this crosstalk leads to qubit error depends strongly on the state of qubit and operation being performed. 

For idle qubits, crosstalk spurs of the type shown here are unlikely to produce qubit-flip transitions, although they will lead to brief changes in the confinement potential determining the sample of nuclear spins that overlap the electron wavefunction. For dynamical decoupling pulse sequences that mitigate slow evolution of the environment \cite{Khodjasteh:2007bu}, the presence of crosstalk spurs between pulses can led to asymmetries in the sequence which produce unwanted qubit phase accumulation \cite{Bluhm}. We note however, that these crosstalk spurs are short on the timescale of typical single-qubit control sequences. Qubits based on the selective addressing of spins using electron spin resonance are also affected by the presence of crosstalk at the level shown here. For instance, EM coupling at resonance frequencies can led to unwanted spin rotation of neighbouring qubits. Strong dc magnetic field gradients between qubits, produced using micro-fabricated magnets \cite{Brunner:2011gma,Pioro} or programmed nuclear fields\cite{Foletti:2009hka}, can alleviate the effect of crosstalk in this regime. 

Crosstalk presents a more significant challenge for two-qubit gates based on capacitively coupled quantum dots or two-electron exchange interaction \cite{Yacoby2qubit,vanWeperen:2011fl}. Here, the presence of unmitigated voltage spurs can lead to over or under rotations of the qubit state-vector, both during idle and active control modes. It also leads to a modulation of the qubit-qubit coupling strength by modifying the position of the electrons and, of concern for exchange based coupling, can modulate the electron wavefunction overlap. In particular, strong exchange coupling requires high bandwidth control in order to produce precise qubit rotations and interactions. This high-speed operation, in combination with the non-linear dependence of exchange coupling with gate voltage, leads to an increased sensitivity to capacitively coupled crosstalk of the type discussed here. For future scaled-up devices with hundreds of gates, the use of multi-layer gate metallisation and ground planes, separated by low loss dielectrics, appears warranted in mitigating EM crosstalk.

In conclusion, we have measured on-chip crosstalk performance of spin qubit devices and developed methods to mitigate crosstalk in multi-qubit architectures. The use of ground guards between on-chip interconnects and miniaturised ohmic contacts has been shown to suppress EM crosstalk to levels below 1\% for nearest neighbour gates. These results indicate that crosstalk can be largely mitigated in present multi-qubit devices and provides  promise for the future scalability of quantum computing architectures based on spins in semiconductors. 

We thank Xanthe Croot, Alice Mahoney, and Andrew Doherty for assistance and discussions and John Reno for providing heterostructure that was a precursor to this work. We acknowledge funding from the U.S. Intelligence Advanced Research Projects Activity (IARPA), through the U.S. Army Research Office and the Australian Research Council Centre of Excellence Scheme (Grant No. EQuS CE110001013).\\

* email: david.reilly@sydney.edu.au

\bibliographystyle{apsrev4-1}

\end{document}